\documentclass[journal,10pt]{IEEEtran}

\ifCLASSINFOpdf
\usepackage[pdftex]{graphicx}

\else

\fi

\usepackage[cmex10]{amsmath}
\usepackage{amssymb}   
\usepackage{amsxtra}
\usepackage{amscd}
\usepackage{amsthm}
\usepackage{textcomp}
\usepackage{graphicx}  
\usepackage{mathtools}
\usepackage{stfloats}
\usepackage{blindtext}
\usepackage{array}  
\makeatletter
\renewcommand*\env@matrix[1][*\c@MaxMatrixCols c]{%
	\hskip -\arraycolsep
	\let\@ifnextchar\new@ifnextchar
	\array{#1}}
\makeatother
\usepackage{multirow}  

\usepackage{amsmath}

\newcommand{\textoverbrace}[2]{%
	\ensuremath{\overbrace{\text{#1}}^{\text{#2}}}%
}

\usepackage{cite}
\usepackage{color,soul} 
\soulregister\cite7
\soulregister\ref7
\soulregister\pageref7

\begin{document}
	\setstcolor{red}
	%
	
	\title{{\LARGE Coordinate Interleaved Orthogonal Design with Media-Based Modulation}}

	\author{Ibrahim~Yildirim,~\IEEEmembership{Student Member,~IEEE}, Ertugrul~Basar,~\IEEEmembership{Senior Member,~IEEE} and\\ Ibrahim Altunbas,~\IEEEmembership{Senior Member,~IEEE} \vspace*{-0.15cm}
		%
		\thanks{This work was supported by the Scientific and Technological Research Council of Turkey (TUBITAK) under Grant 117E869.}
		\thanks{I. Yildirim and I. Altunbas are with Istanbul Technical University, Istanbul, Turkey. (e-mail: yildirimib@itu.edu.tr, ibraltunbas@itu.edu.tr)
			
			E. Basar is with Koç University, Istanbul, Turkey. (e-mail: ebasar@ku.edu.tr)}
	}

	\markboth{ }{}

	
	\maketitle

	\begin{abstract}
	In this work, we propose a new multiple-input multiple-output (MIMO) concept, which is called coordinate interleaved orthogonal design with media-based modulation (CIOD-MBM). The proposed two novel CIOD-MBM schemes provide improved data rates as well as diversity gain while enabling hardware simplicity using a single radio frequency (RF) chain. Moreover, using the equivalent channel model, a reduced complexity can be obtained for maximum likelihood (ML) detection of the proposed system. Using computer simulations, it is shown that CIOD-MBM schemes provide remarkably better performance against the conventional MBM and CIOD systems.

	\end{abstract}

	\begin{IEEEkeywords}
		Space-time coding, coordinate interleaved orthogonal design, index modulation, media-based modulation.
	\end{IEEEkeywords}

	%
	\IEEEpeerreviewmaketitle
	
	
	\section{Introduction}

	\IEEEPARstart{W}{ith} the proliferation of applications using wireless communication systems in recent years, a huge increase in the number of users and data traffic has emerged. It is envisaged that next-generation wireless communication applications such as Internet-of-Things, vehicle-to-everything communications and molecular communications will be introduced into daily life in the near future \cite{5G}. Despite the increasing demand from day to day, the available resources remain limited when considering the frequency spectrum and physical hardware implementations. In next-generation communication systems, under such resource constraints, it is aimed to develop methods with high band efficiency, channel capacity and energy efficiency. All around the world, researchers have eagerly worked to meet these growing demands. As a result of these efforts, many versatile and highly successful methods have been put forward.  
	
	Index modulation (IM) schemes, which achieve high data rates while reducing hardware costs in the transmitter, have recently received remarkable attention. Spatial modulation (SM), the well-known member of the IM family, conveys information by transmit antenna indices as well as conventional phase shift keying (PSK) or quadrature amplitude modulation (QAM) methods \cite{SM_optimal}. When high data rates are desired in SM systems, it is required to make giant increases in the number of transmitting antennas. From this point of view, the media-based modulation (MBM) scheme, which is proposed as an avant-garde digital transmission method for rich scattering environments, is considered as an alternative to SM \cite{Khandani_conf1}. In MBM systems, the current distributions of the reconfigurable antennas are changed by means of various elements, such as PIN diodes, varactors, micro-electro-mechanical switches, and so on,  hence, different channel fading realizations, which is called as channel states, are obtained. Using the on/off status of the parasitic elements called radio frequency (RF) mirrors, different radiation patterns are acquired by varying current characteristics.  If these antenna radiation patterns are sufficiently uncorrelated, distinguishable channel fading realizations can be obtained and used as an additional information source in a rich scattering environment.

	Since a single transmit antenna will be active in SM and MBM systems during the transmission, inter-channel interference (ICI) problem is eliminated. On the other hand, the transmission via a single active antenna means that a single RF chain is used in the transmitter part, which greatly reduces the hardware cost. Furthermore, unlike SM, since the spectral efficiency in the MBM varies linearly proportional to the number of RF mirrors, it is sufficient to increase the number of RF mirrors in a bid to achieve high data rates.
	
	Existing multiple-input multiple-output (MIMO) techniques such as SM,  quadrature SM (QSM) and generalized SM (GSM) can be integrated with MBM to achieve promising results. In \cite{MBM_TVT}, GSM is adopted to MBM in order to enhance spectral efficiency and error performance. Combining QSM and MBM innovatively, promising results are obtained by quadrature channel modulation (QCM) \cite{QCM}. Space-time coding concept also can be merged with MBM to increase the diversity gain. To this end, space-time channel modulation (STCM) method is proposed in \cite{STCM}, which combines Alamouti's famous space-time block code (STBC) with MBM. Additionally, considering the flexibility of IM techniques, time-indexed based SM, MBM, and SM-MBM schemes are proposed in \cite{Multi_IM}. In recent past, since MBM systems are prone to increase physical layer security, various secret transmission scenarios are proposed to enable high secrecy capacity using appropriate designs \cite{MBM_sec,MBM_sec2}. Differential MBM (DMBM) \cite{Diff_MBM} concept has alleviated the requirement of channel sounding when it is hard to get accurate channel state information (CSI) at the destination. In \cite{STMBM}, a comprehensive frame for space-time coded MBM systems is presented and STBC-based MBM scheme, which uses a single RF chain, is proposed and miscellaneous diversity gains are obtained. In order to overcome fractional bit problems in IM systems, authors propose fractional MBM with golden angle modulation (GAM-MBM) in \cite{GAM_MBM} by making use of the symmetrical structure of GAM.
	
	In this paper, we introduce coordinate interleaved orthogonal design MBM (CIOD-MBM) concept, which provides diversity gain as well as increased spectral efficiency while assuring hardware simplicity using a single RF chain.  The focus of this work is to bring the desirable benefits of CIOD, such as enabling low complexity maximum likelihood (ML) detection opportunity and simple transmitter structure, into the area of interest of MBM. In this context, two promising CIOD-MBM schemes, which use a single active antenna during the transmission, are presented. Furthermore, the theoretical average bit error probability (ABEP) of the proposed CIOD-MBM schemes are derived. Afterwards, the supremacy of the proposed CIOD-MBM concept is demonstrated via computer simulations in terms of error performance and obtained numerical results are verified with theoretical results.


	\section{Coordinate Interleave Orthogonal Designs for Media-Based Modulation}
	
		\begin{figure*}[t]
		\begin{center}
			{\includegraphics[scale=1]{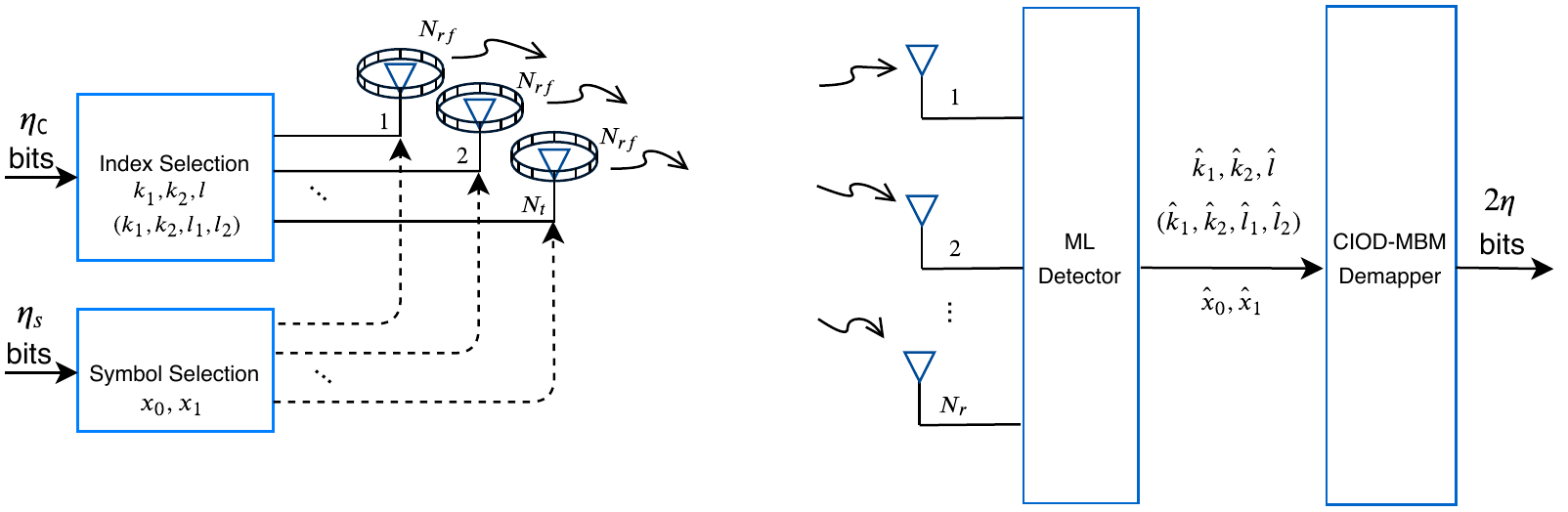}}
			\vspace*{-0.3cm}
			\caption{Block diagram of $N_{r}\times N_{t}$ MIMO system which is employed by CIOD-MBM schemes.}
			\vspace*{-0.3cm}
		\end{center}
	\end{figure*}
	
	In this section, we introduce the system model and the working principle of CIOD-MBM schemes under quasi-static flat Rayleigh fading channel. $N_r \times N_t $ MIMO transmission scenario is considered by assuming each transmit antenna has $N_{rf}$ RF mirrors. In the proposed systems, CIOD scheme contributes to MBM transmissions as an additional source of information.

	The essential principle of CIOD is to achieve diversity gain by transmitting symbols through different channels in two different time slots. Due to the transmission using a single active antenna, reduced hardware cost is obtained keeping only one RF chain. Unlike traditional STBCs, signal constellation must be rotated at a specific angle to ensure the full diversity of CIODs \cite{CIOD}. The CIOD signal matrix for the system with two transmit antennas is as follows
	\begin{equation}
	\begin{bmatrix}[cc] \tilde{s_0} & 0   \\	0 & \tilde{s_1}  \\ \end{bmatrix}
	\end{equation}
	where $\tilde{s_0}=x_0^\Re+jx_1^\Im $ and $\tilde{s_1}=x_1^\Re+jx_0^\Im$ are transmitted symbols, which are obtained by partitioning $x_i$ ($i \in \left\lbrace 0,1 \right\rbrace $) symbol into its real ($x_i^\Re$) and imaginary ($x_i^\Im$) part. Here, $x_0$ and $x_1$ are selected from a signal constellation formed by rotating the ordinary $M$-ary signal constellation at a certain angle. For instance, the optimum rotation angle in the quadrature PSK (QPSK) constellation is 13.2885° for the conventional CIOD in (1) \cite{CIOD2}.  
	
	Considering peculiarities of CIOD, two novel CIOD-MBM schemes are designed for a given MIMO system. The generalized baseband received signal model of the CIOD-MBM schemes can be expressed as
	\begin{equation}\label{model}
	\mathbf{Y}=\mathbf{H X }+\mathbf{ N}
	\end{equation}
	where $\mathbf{Y} \in \mathbb{C}^{N_r \times 2}$ is matrix of the received signals  in two different time slots, $\mathbf{H} \in \mathbb{C}^{N_r \times N_t2^{N_{rf}}}$  is the channel matrix, and $\mathbf{N} \in \mathbb{C}^{N_r \times 2}$ is the noise sample matrix, respectively. At the same time, $\mathbf{X}\in \mathbb{C}^{N_t2^{N_{rf}}\times 2}$ represents the CIOD matrix, which is designed for two different MBM schemes. The elements of $\mathbf{H}$ and $\mathbf{N}$ follow zero mean independent and identically complex Gaussian distribution with variances $1$ and $N_0$, respectively.

	\subsection{CIOD-MBM I Scheme}
	The block diagram of CIOD-MBM I scheme is given in Fig. 1, where $\eta_c=\log_2(N_t/2)+N_{rf}$ and $\eta_s=2\log_2(M)$ respectively denote the number of index and symbol bits sent in two slots. The spectral efficieny of CIOD-MBM I scheme is calculated as
	\begin{equation}
	\eta = \frac{\log_2(N_t/2)+N_{rf}+2\log_2(M)}{2}  \text{\quad bit/s/Hz}.
	\end{equation}
	
	Before determining the active channel state, available antennas are divided into two groups. Then, using the first $\log_2(N_t/2)$ bits, active antenna index $k_1$ is determined for the first time slot. The group, to which the first active antenna belongs, is omitted in the second selection. For the determined antenna, active channel state $l$ is chosen depending on $N_{rf}$ bits. In the second time slot, second active antenna index $k_2$ is determined as $k_2=N_t/2+k_1$. Then, $l$th channel state is also selected in the second time slot as similar as first time slot. After all, transmitted symbols $x_0$ and $x_1$ are chosen by the aid of last $2\log_2(M)$ bits. Following these procedures, the transmission matrix $\mathbf{X}$ is constructed as
	\newcommand\undermat[2]{
		\makebox[0pt][l]{$\smash{\underbrace{\phantom{%
						\begin{matrix}#2\end{matrix}}}_{\text{$#1$}}}$}#2}
		\begin{equation}
{\footnotesize \mathbf{X}=\begin{bmatrix}[c|c|c|c|c] \dots & \dots  \textoverbrace{$ \tilde{s_0}  $}{$l$\text{th state}}   \dots & \dots & 0 \quad \dots \quad 0 &\dots \\ 
\dots	& \undermat{k_1\text{th antenna}}{\hphantom{0}{\dots \quad 0 \quad \dots}}  & \dots & 	\undermat{k_2\text{th antenna}}{\hphantom{0}{\dots \quad {\textoverbrace{$ \tilde{s_1}  $}{$l$\text{th state}}} \quad \dots}} & \dots  \\ \\ \end{bmatrix}^T}
\end{equation}
	\\
	where $(.)^T$ denotes matrix transpose, and possible channel realizations and time slots are represented by rows and columns, respectively.

	The CIOD-MBM I scheme provides full diversity when the minimum determinant of the distance matrix is maximized \cite{Jafarkhani}. Under the full rank condition, the minimum coding gain distance is expressed as	
	\begin{equation}\delta_{\min}=\underset{\mathbf{X},\hat{\mathbf{X}},\mathbf{X}\neq \hat{\mathbf{X}}}{\mathop{\min}}\, \det[(\mathbf{X}-\hat{\mathbf{X}})(\mathbf{X}-\hat{\mathbf{X}})^H].
	\end{equation}

	In CIOD-MBM I, the rotation angle, which guarantees full diversity, is calculated as $ 13.2885 $° for QPSK. 
	If the system with $N_t=4$, $N_{rf}=1$ and $M=4$ (QPSK) is considered to exemplify the working mechanism of CIOD-MBM I, spectral efficiency ($\eta$) is calculated as $3$ bit/s/Hz. In two time slots, $2\eta=6$ incoming bits are processed as follows:
	
	\begin{equation}
	{\footnotesize \mathbf{b}=[\,\underbrace{1\ 0}_{\mathop{\log_2(N_t/2)+N_{rf}}}\,\underbrace{1\ 1}_{\mathop{\log _2(M)}}\,\underbrace{1\ 0}_{\mathop{\log _2({M})}}]}.  
	\end{equation} 
	Using the mapping rule of the CIOD-MBM as shown in Table I, first $\log_2(N_t/2)+N_{rf}=2$ bits assign the active antenna index set $\left\lbrace k_1,k_2\right\rbrace$ and channel state $l$, respectively. After the two active channel states are separately obtained, the next $[1\ 1]$ bits modulate to determine rotated QPSK symbol $x_0=0.2299-j0.9732$. Thereafter, last two bits $[1\ 0]$ specify the $x_1=-0.9732 - j0.2299$ from rotated QPSK symbol set. Finally, CIOD-MBM I matrix  is obtained as
		\begin{equation}
	      \mathbf{ X}=\begin{bmatrix}[cc|cc|cc|cc] 0 & 0 & x_0^\Re+jx_1^\Im  & 0 & 0 & 0 & 0 & 0  \\	0 & 0 & 0 & 0 & 0 & 0 & x_1^\Re+jx_0^\Im  &  0 \\ \end{bmatrix}^T.
	      	\end{equation}

	\begin{table}[t]
		\caption{CIOD-MBM index mapping rule for $N_t=4$ and $N_{rf}=1$} 
		\centering 
		
		\begin{tabular}{c c c c} 
			\hline\hline 
			\rule{0pt}{12pt}Bits & $\left\lbrace k_1,k_2\right\rbrace$ &  $ l $& $\mathbf{X}^T$\\ [0.5ex] 
			\hline 
			& & &\\
			\rule{0pt}{5pt}00 & $\left\lbrace 1,3 \right\rbrace$ & $ 1$&   $\begin{bmatrix}[cc|cc|cc|cc] \tilde{s_0} & 0 & 0 & 0 & 0 & 0 & 0 & 0  \\	0 & 0 & 0 & 0 & \tilde{s_1} & 0 & 0 &  0 \\ \end{bmatrix}$\\
			\rule{0pt}{16pt}01 & $\left\lbrace 1,3 \right\rbrace$ & $2$& $\begin{bmatrix}[cc|cc|cc|cc] 0 & \tilde{s_0} & 0 & 0 & 0 & 0 & 0 & 0  \\	0 & 0 & 0 & 0 & 0 & \tilde{s_1} & 0 &  0 \\ \end{bmatrix}$\\
			\rule{0pt}{16pt}10 & $\left\lbrace 2,4 \right\rbrace$ & $1$& $\begin{bmatrix}[cc|cc|cc|cc] 0 & 0 & \tilde{s_0} & 0 & 0 & 0 & 0 & 0  \\	0 & 0 & 0 & 0 & 0 & 0 & \tilde{s_1} &  0 \\ \end{bmatrix}$\\
			\rule{0pt}{16pt}11 & $\left\lbrace 2,4 \right\rbrace$ & $2$& $\begin{bmatrix}[cc|cc|cc|cc] 0 & 0 & 0 & \tilde{s_0} & 0 & 0 & 0 & 0  \\	0 & 0 & 0 & 0 & 0 & 0 & 0 &  \tilde{s_1} \\ \end{bmatrix}$\\
			[1ex] 
			\hline 
		\end{tabular}
		\label{table:nonlin} 
		\vspace*{-0.5cm}
	\end{table}
	
		\subsection{CIOD-MBM II}    
The main motivation of this system is to increase the spectral efficiency of the CIOD-MBM I scheme while maintaining diversity gain and other inherent advantages. The block diagram of CIOD-MBM II can be seen from Fig. 1, which the terms in parentheses represent active antenna and channel state indices of CIOD-MBM II, for $\eta_c=N_{rf}-1+2\log_2(N_t)$ and $\eta_s=2\log_2(M)$. In the CIOD-MBM II scheme, which is inspired by QSM \cite{QSM} and QCM \cite{QCM}, active transmit antennas are individually selected for real and imaginary parts of $\tilde{s_0} $ and $\tilde{s_1} $ symbols. The spectral efficiency of the CIOD-MBM II is calculated as
	\begin{equation}
\eta = \frac{N_{rf}-1+2\log_2(N_t)+2\log_2(M)}{2}  \text{\quad bit/s/Hz}.
\end{equation}

In order to guarantee the utilization of a single RF chain at a single time slot, a total of $2^{N_{rf}}$ channel states are divided into two groups similar to the CIOD-MBM I. The first $ 2N_{rf}-1 $ bits decide the active channel state $ l_1 $ for the initial time slot. After determining
the group, to which $ l_1 $ belongs, active second channel
state $ l_2 $ is determined as $l_2 =2^{N_{rf}}/2+l_1 $ . Using $2\log_2(N_t)$ bits, active antennas $k_1$ and $k_2$ are selected for real and imaginary parts of transmitted symbols, respectively. Then, the last $2\log_{2}(M)$ bits are used to specify the modulated symbols $x_0$ and $x_1$ from rotated $M$-ary constellations. Generalized transmission matrix is given as
\begin{align}
{\footnotesize \mathbf{X}=\begin{bmatrix}[c|c|c|c|c] \dots & \dots\textoverbrace{$ x_0^\Re  $}{$k_{1}$}   \dots  \textoverbrace{$j x_1^\Im  $}{$k_{2}$}  \dots & \dots & \dots \quad 0 \quad \dots &\dots \\ 
\dots	& \undermat{l_1\text{th state}}{\hphantom{0}{\dots \quad 0 \quad \dots}} & \dots  &\undermat{l_2\text{th state}}{\hphantom{0}{\dots\textoverbrace{$ x_1^\Re  $}{$k_{1}$}   \dots  \textoverbrace{$j x_0^\Im  $}{$k_{2}$}  \dots}}& \dots \\ \end{bmatrix}^T}\\ \nonumber
\end{align}
where possible channel realizations and time slots are represented by rows and columns, respectively.

	\begin{table}[t]
	\caption{CIOD-MBM II index mapping rule for $N_t=2$ and $N_{rf}=2$} 
	\centering 
	
	\begin{tabular}{c c  c} 
		\hline\hline 
		\rule{0pt}{12pt} Bits & $\left\lbrace l_1,l_2 \right\rbrace$-$\left\lbrace k_1,k_2\right\rbrace$& $\mathbf{X}^T$\\ [0.5ex] 
		\hline 
		&  &\\
		\rule{0pt}{5pt}000 &$\left\lbrace 1,3 \right\rbrace$ - $\left\lbrace 1,1 \right\rbrace$&  $\begin{bmatrix}[cc|cc|cc|cc] x_0^\Re+jx_1^\Im & 0 &  0 & 0 & 0 & 0 & 0 & 0  \\	0 & 0 & 0 & 0 & x_1^\Re+jx_0^\Im & 0 & 0  &  0 \\ \end{bmatrix}$ \\
		\rule{0pt}{16pt}001 &$\left\lbrace 1,3 \right\rbrace$ - $\left\lbrace 1,2 \right\rbrace$&  $\begin{bmatrix}[cc|cc|cc|cc] x_0^\Re+jx_1^\Im & 0 &  0 & 0 & 0 & 0 & 0 & 0  \\	0 & 0 & 0 & 0 & x_1^\Re & jx_0^\Im & 0  &  0 \\ \end{bmatrix}$ \\
		\rule{0pt}{16pt}010 &$\left\lbrace 1,3 \right\rbrace$ - $\left\lbrace 2,1 \right\rbrace$&  $\begin{bmatrix}[cc|cc|cc|cc] x_0^\Re & jx_1^\Im &  0 & 0 & 0 & 0 & 0 & 0  \\	0 & 0 & 0 & 0 & x_1^\Re+jx_0^\Im & 0 & 0  &  0 \\ \end{bmatrix}$\\
		\rule{0pt}{16pt}011 &$\left\lbrace 1,3 \right\rbrace$ - $\left\lbrace 2,2 \right\rbrace$&  $\begin{bmatrix}[cc|cc|cc|cc] x_0^\Re & jx_1^\Im &  0 & 0 & 0 & 0 & 0 & 0  \\	0 & 0 & 0 & 0 & x_1^\Re & jx_0^\Im & 0  &  0 \\ \end{bmatrix}$  \\
		\rule{0pt}{16pt}100 &$\left\lbrace 2,4 \right\rbrace$ - $\left\lbrace 1,1 \right\rbrace$&  $\begin{bmatrix}[cc|cc|cc|cc] 0 & 0 &  x_0^\Re+jx_1^\Im & 0 & 0 & 0 & 0 & 0  \\	0 & 0 & 0 & 0 & 0 & 0 & x_1^\Re+jx_0^\Im  &  0 \\ \end{bmatrix}$  \\
		\rule{0pt}{16pt}101 &$\left\lbrace 2,4 \right\rbrace$ - $\left\lbrace 1,2 \right\rbrace$&  $\begin{bmatrix}[cc|cc|cc|cc] 0 & 0 &  x_0^\Re+jx_1^\Im & 0 & 0 & 0 & 0 & 0  \\	0 & 0 & 0 & 0 & 0 & 0 & x_1^\Re  &  jx_0^\Im \\ \end{bmatrix}$  \\
		\rule{0pt}{16pt}110 &$\left\lbrace 2,4 \right\rbrace$ - $\left\lbrace 2,1 \right\rbrace$&  $\begin{bmatrix}[cc|cc|cc|cc] 0 & 0 &  x_0^\Re & jx_1^\Im & 0 & 0 & 0 & 0  \\	0 & 0 & 0 & 0 & 0 & 0 & x_1^\Re+jx_0^\Im  &  0 \\ \end{bmatrix}$  \\
		\rule{0pt}{16pt}111 &$\left\lbrace 2,4 \right\rbrace$ - $\left\lbrace 2,2 \right\rbrace$&  $\begin{bmatrix}[cc|cc|cc|cc] 0 & 0 &  x_0^\Re & jx_1^\Im & 0 & 0 & 0 & 0  \\	0 & 0 & 0 & 0 & 0 & 0 & x_1^\Re  &  jx_0^\Im \\ \end{bmatrix}$  \\
		[1ex] 
		\hline 
	\end{tabular}
	\label{table:nonlin} 
	\vspace*{-0.5cm}
\end{table}

The system with $N_t=2$, $N_{rf}=2$ and $M=4$ is explained by the following example. In this scenario, the spectral efficiency is calculated as $\eta=3.5$ bit/s/Hz. $ 30 $° is the optimum angle that ensures full diversity and maximizes the $\delta_{min}$ for QPSK in CIOD-MBM II while $ 8.6 $° is the optimum angle for $16$-QAM. For two time slots, it is assumed that incoming bits are arranged as follows:
      	\begin{equation}
      {\footnotesize \mathbf{b}=[\,\underbrace{1\ 0\ 1}_{\mathop{N_{rf}-1+2\log_2(N_t)}}\,\underbrace{0\ 1}_{\mathop{\log _2(M)}}\,\underbrace{1\ 1}_{\mathop{\log _2({M})}}]}.  
      \end{equation}
With the aid of Table II, active antenna set $\left\lbrace k_1,k_2\right\rbrace$ and channel index set $\left\lbrace l_1,l_2\right\rbrace$ are determined by the first $\eta_c=3$ index bits. Then, using $\eta_s$ symbol bits, $x_0$ and $x_1$ are chosen from the rotated constellation as $-0.5+j0.866$ and $0.5-j0.866$, respectively. In the final stage, transmission matrix is constructed as
\begin{equation}
 \mathbf{ X}=\begin{bmatrix}[cc|cc|cc|cc] 0 & 0 &  -0.5-j0.866 & 0 & 0 & 0 & 0 & 0  \\	0 & 0 & 0 & 0 & 0 & 0 & 0.5  &  j0.866 \\ \end{bmatrix}^T.
\end{equation}

	\subsection{ML Detection for Proposed CIOD-MBM Schemes}
	As mentioned previous sections, one of the major advantages of CIOD is its symbol-by-symbol decoding possibility while keeping ML complexity remarkably reduced. In this sense, in the proposed CIOD-MBM I scheme, the equivalent system model of (2) is obtained by
	\begin{equation}
		\mathbf{y}_{eq}=\mathbf{H}_{eq}\mathbf{x}_{eq}+\mathbf{n}_{eq}
	\end{equation}
	where $\mathbf{y}_{eq} \in \mathbb{C}^{4N_r \times 1} $, $\mathbf{x}_{eq}=[x_0^\Re \ x_0^\Im \ x_1^\Re \ x_1^\Im]^T$ and $\mathbf{n}_{eq} \in \mathbb{C}^{4N_r \times 1}$ represent the equivalent received signal vector,  the equivalent transmitted symbols and the equivalent noise vector, respectively. The construction of generalized $\mathbf{y}_{eq}$ is shown by	
	\begin{equation}
\mathbf{y}_{eq} =[y_{1,1}^\Re \ y_{2,1}^\Re \dots y_{N_r,1}^\Re \ y_{1,1}^\Im \dots y_{N_r,1}^\Im \dots y_{1,2}^\Re \dots y_{N_r,2}^\Im]^T
	\end{equation}
	where $y_{r,t}$ stands for received signal at $r$th receive antenna in $t$th time slot.
	
	\begin{figure*}[!t]
		\normalsize
		\begin{equation}
		\label{Heq_1}
		\mathbf{H}_{eq}=\begin{bmatrix}[cccc cccc cccc c] h_{1,m_1}^\Re  & h_{2,m_1}^\Re  & \dots & h_{N_r,m_1}^\Re &  h_{1,m_1}^\Im & \dots & h_{N_r,m_1}^\Im & 0 & \dots & 0 &0 & \dots & 0  \\	
		0  & 0  & \dots & 0 &  0& \dots & 0 & h_{1,m_2}^\Re & \dots & h_{N_r,m_2}^\Re & h_{1,m_2}^\Im & \dots & h_{N_r,m_2}^\Im  \\
		0  & 0  & \dots & 0 &  0& \dots & 0 & -h_{1,m_2}^\Im & \dots & -h_{N_r,m_2}^\Im & h_{1,m_2}^\Re & \dots & h_{N_r,m_2}^\Re  \\
		-h_{1,m_1}^\Im  & -h_{2,m_1}^\Im  & \dots & -h_{N_r,m_1}^\Im &  h_{1,m_1}^\Re & \dots & h_{N_r,m_1}^\Re & 0 & \dots & 0 &0 & \dots & 0  \\  \end{bmatrix}^T
		\end{equation}
		
		\hrulefill
		\vspace*{-0.5cm}
	\end{figure*}
	
	Furthermore, $\mathbf{H}_{eq} \in \mathbb{C}^{4N_r \times 4}$ denotes the equivalent channel matrix, which is specified in (\ref{Heq_1}) for CIOD-MBM I. In this notation, $h_{r,m_i}$ ($i \in \left\lbrace 1,2 \right\rbrace $) represents channel fading coefficients between the selected active channel state ($m_i$) and $r$th receive antennas, where $m_i$ is chosen from the possible $N_t2^{N_rf}$ channel realizations depending on $k$ and $l$ sets. Because of the orthogonality of columns, the equivalent channel matrix is decomposed as 	
	$  \mathbf{H}_{eq}=[\mathcal{H}_1 \ \mathcal{H}_2] $
	where $\mathcal{H}_1  \in \mathbb{C}^{4N_r \times 2} $ and $\mathcal{H}_2\in \mathbb{C}^{4N_r \times 2} $ represent first two and last two columns of $\mathbf{H}_{eq}$. It can be easily observed that $\mathcal{H}_1$ and $\mathcal{H}_2$ are orthogonal each other, therefore, $x_0$ and $x_1$ are individually decoded for CIOD-MBM I. 
	
	As is common, considering the perfect CSI in CIOD-MBM I, ML detection is employed at the receiver by
	\begin{equation}
	\label{ML_rule}
	[\hat{x_0},\hat{x_1},\hat{k}_1,\hat{k}_2,\hat{l}]=\arg\underset{x_0,x_1,k_1,k_2,l}{\mathop{\min}}\,{{\left\| \mathbf{y}-\mathbf{HX} \right\|}^{2}}
		\end{equation}
	where $\Arrowvert.\Arrowvert$ represents the Euclidean norm. Similarly, the detection of the index bits is performed by minimization over $k_1,k_2,l_1,l_2$  indices in (\ref{ML_rule}) instead of $k_1,k_2,l$ in CIOD-MBM II scheme.
	
	In CIOD-MBM I, considering all possible channel realizations, minimum ML decision metrics of $x_0$ and $x_1$ are separately obtained by
		\begin{align}d_0^{(k_1,l)}=\underset{x_0}{\mathop{\min}}\,{{\left\| \mathbf{y}-\mathcal{H}_1[x_0^\Re \ x_0^\Im]^T  \right\|}^{2}}, \nonumber \\
	d_1^{(k_2,l)}=\underset{x_1}{\mathop{\min}}\,{{\left\| \mathbf{y}-\mathcal{H}_2[x_1^\Re \ x_1^\Im]^T  \right\|}^{2}}. 
	\end{align}
	Thereafter, the minimum ML decision metric for a corresponding antenna and antenna state indices $(k_1,k_2,l)$ is calculated by $d^{(k_1,k_2,l)}=d_0^{(k_1,l)}+d_1^{(k_2,l)}$. After obtaining all $d^{(k_1,k_2,l)}$, the most likely antenna and antenna state indices combination is determined by $(\hat{k}_1,\hat{k}_2,\hat{l})=\arg \min_{(k_1,k_2,l)} d^{(k_1,k_2,l)} $. Using the most likely channel realizations, the transmitted data symbols are estimated by $(\hat{x}_0,\hat{x}_1)=(x_0^{(k_1,l)},x_1^{(k_2,l)})$.

	After the demapping process, decoded bits are obtained at the receiver. Thanks to the equivalent model, the number of calculated metrics is reduced $2^{\eta_c}M^2$ to $2^{\eta_c}2M$ for CIOD-MBM I scheme. Due to the non-ortogonality of equivalent channel matrix for CIOD-MBM II, brute-force ML detection is performed as in (\ref{ML_rule}) with a complexity order of $2^{\eta_c}M^2$.

	\section{Performance Analysis}
	By the aid of union bounding technique, bit error performance of CIOD-MBM schemes are obtained \cite{Simon}. Assuming that $\mathbf{X}$ is transmitted and incorrectly detected as $\hat{\mathbf{X}}$, the ABEP is approximately calculated by
	\begin{equation}{{P}_{b}}\approx \frac{1}{2\eta {{2}^{2\eta }}}\sum\nolimits_{\mathbf{X}}{\sum\nolimits_{\mathbf{\hat{X}}}}{P(\mathbf{X}\to \mathbf{\hat{X}})}\,e(\mathbf{X},\mathbf{\hat{X}})
	\end{equation}
	where pairwise bit error probability (PEP) and the number of bit errors are represented by $P(\mathbf{X}\to \mathbf{\hat{X}})$ and $e(\mathbf{X},\mathbf{\hat{X}})$ respectively.
	
	The conditional PEP (CPEP) for the CIOD-MBM schemes can be obtained as
	\begin{align}
	{\small P(\mathbf{X}\! \to\! \mathbf{\hat{X}}|\mathbf{H})\!=\!Q\!\left(\! \sqrt{\!{{\Delta}}/{(2{N_0})}} \right)}
	\end{align}
	where $\Delta={\small{\left\| \mathbf{H}(\mathbf{X}-\mathbf{\hat{X}}) \right\|}^{2}_\text{F}} $, $\left\|.\right\|^{2}_\text{F}$ is the Frobenius norm and $Q(x)$ is the well-known $Q$-function, which is also represented as $Q(x)=1/\pi\int_{0}^{\pi/2}e^{-x^2/\sin^2\theta}d\theta$. By means of this alternative version of the $Q$-function and the moment generating function of  ${ {\left\| \mathbf{H}(\mathbf{X}-\mathbf{\hat{X}}) \right\|}^{2}_\text{F} }$ \cite{turin}, the unconditional PEP is given by
		\begin{align} P(\mathbf{X}\to& \mathbf{\hat{X}})=\nonumber \\
	=&\frac{1}{\pi }{{\int\nolimits_{0}^{\frac{\pi }{2}}{\,\left( \frac{{{\sin }^{2}}\theta }{{{\sin }^{2}}\theta +\frac{\lambda_1}{4{{N}_{0}}}} \right)}}^{{{N}_{r}}} {\,\left( \frac{{{\sin }^{2}}\theta }{{{\sin }^{2}}\theta +\frac{\lambda_2}{4{{N}_{0}}}} \right)}^{N_r} }\,d\theta
	\end{align}

	where $\lambda_d$, $d \in \left\lbrace 1,2 \right\rbrace$ stand for the eigenvalues of the $(\mathbf{X}-\mathbf{\hat{X}})(\mathbf{X}-\mathbf{\hat{X}})^H$ matrix, which is not equal to zero.
	
%

	\section{Simulation Results}
	In order to evaluate the error performance of the CIOD-MBM schemes, BER results are produced using computer simulations under varying system parameters. Considering conventional MBM \cite{Khandani_conf1}, CIOD \cite{CIOD} and CIOD-SM \cite{CIOD_SM} as reference systems, BER results are obtained with respect to $E_b/N_0$, where $E_b$ is the average transmitted energy per bit.

	In Fig. 2, theoretical ABEP results of both CIOD-MBM schemes are obtained to compare the BER results obtained by computer simulations.As can be clearly seen, the theoretical curves are perfectly matched with the computer simulation results at moderate and high $E_b/N_0$ values. At the same time, the proposed CIOD-MBM schemes offer a flexible design opportunity for varying $N_t$, $N_{rf}$ and $N_r$ values. As expected, the error performance deteriorates as the number of possible channel realizations increases when the same $M$-ary constellation is used. For increasing $\eta$ values, CIOD-MBM II scheme provides better error performance than CIOD-MBM I scheme as clearly seen in Fig. 2.
	
	\begin{figure}[!t]
		\begin{center}\resizebox*{7.5cm}{6cm}{\includegraphics{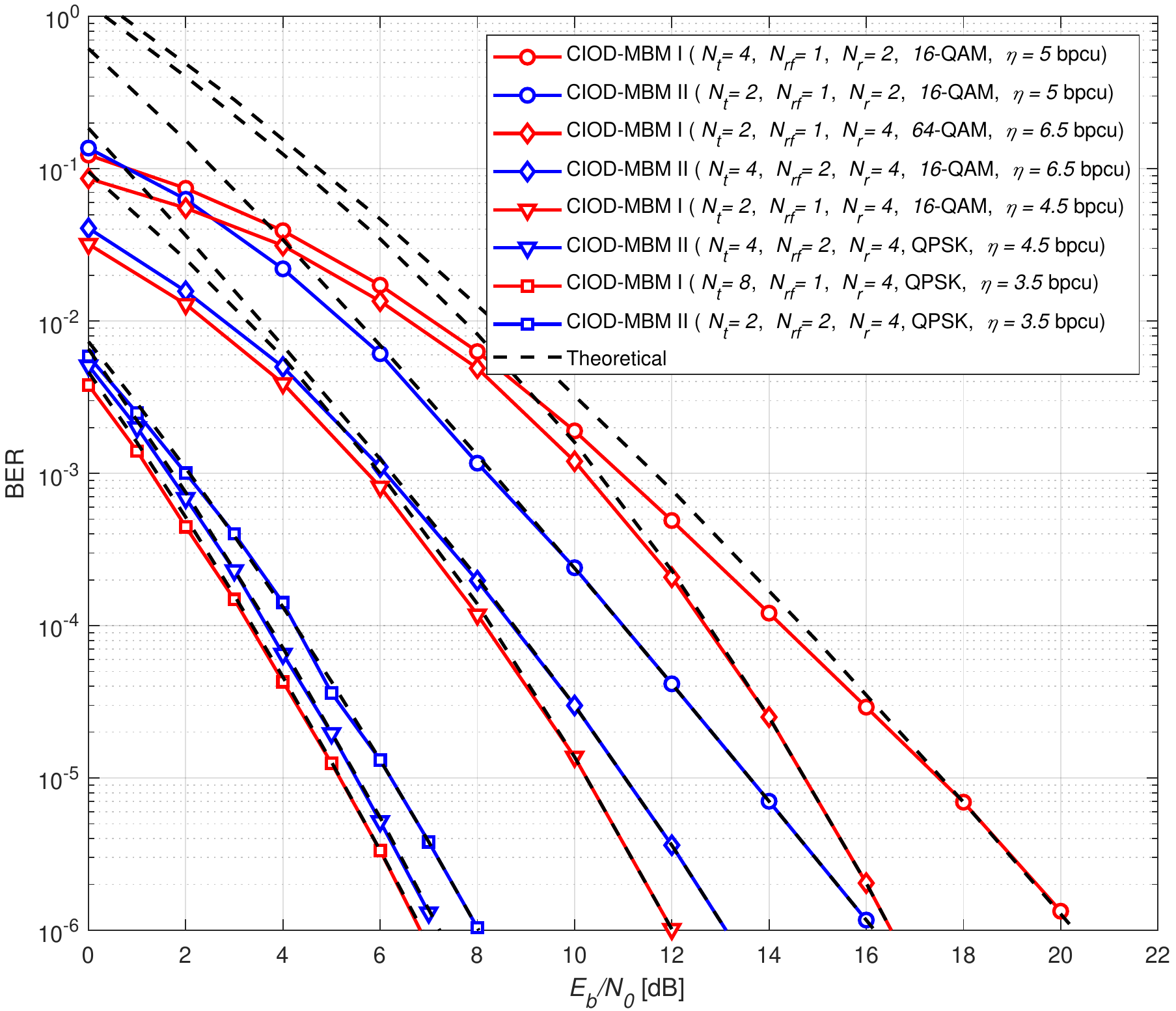}}
			\vspace*{-0.0cm}\caption{Theoretical ABEP and computer simulated BER of the CIOD-MBM schemes under various $N_t$, $N_{rf}$, $N_r$ and $M$ values.}\vspace*{-0.5cm}
		\end{center}
	\end{figure}
	
	Fig. 3 depicts the BER of the proposed schemes against the BER of conventional CIOD, CIOD-SM and MBM for $\eta=4$ bit/s/Hz. Here, degree of SM (DoSM) scheme, which is defined in \cite{CIOD_SM}, is considered as CIOD-SM. It should be noted that all reference schemes use a single RF chain during the transmission for a fair comparison. As can be clearly seen, CIOD-MBM I scheme provides around 5 dB supremacy over CIOD and approximately 10 dB over CIOD-SM and MBM schemes. At the same time, CIOD-MBM II scheme provides similar error performance with CIOD-MBM scheme for $\eta=4$ bit/s/Hz while providing better error performance for higher spectral efficiency values as in Fig. 2.

	\section{Conclusions}
	In this paper, CIOD-MBM concept has been introduced to increase the spectral efficiency of conventional CIOD and MBM schemes by using CIOD as an additional information source as well as transmit antennas, channel states, and amplitude/phase modulation. One of the most weighty benefits of the CIOD-MBM schemes, which promise high data rates by using a low number of transmit antennas, is enabling reduced receiver complexity by symbol-by-symbol detection. As a result of theoretical ABEP and computer simulations, it has been shown that proposed CIOD-MBM schemes provide superiority to reference systems in terms of error performance.

	\begin{figure}[t]
	\begin{center}\resizebox*{7.5cm}{6cm}{\includegraphics{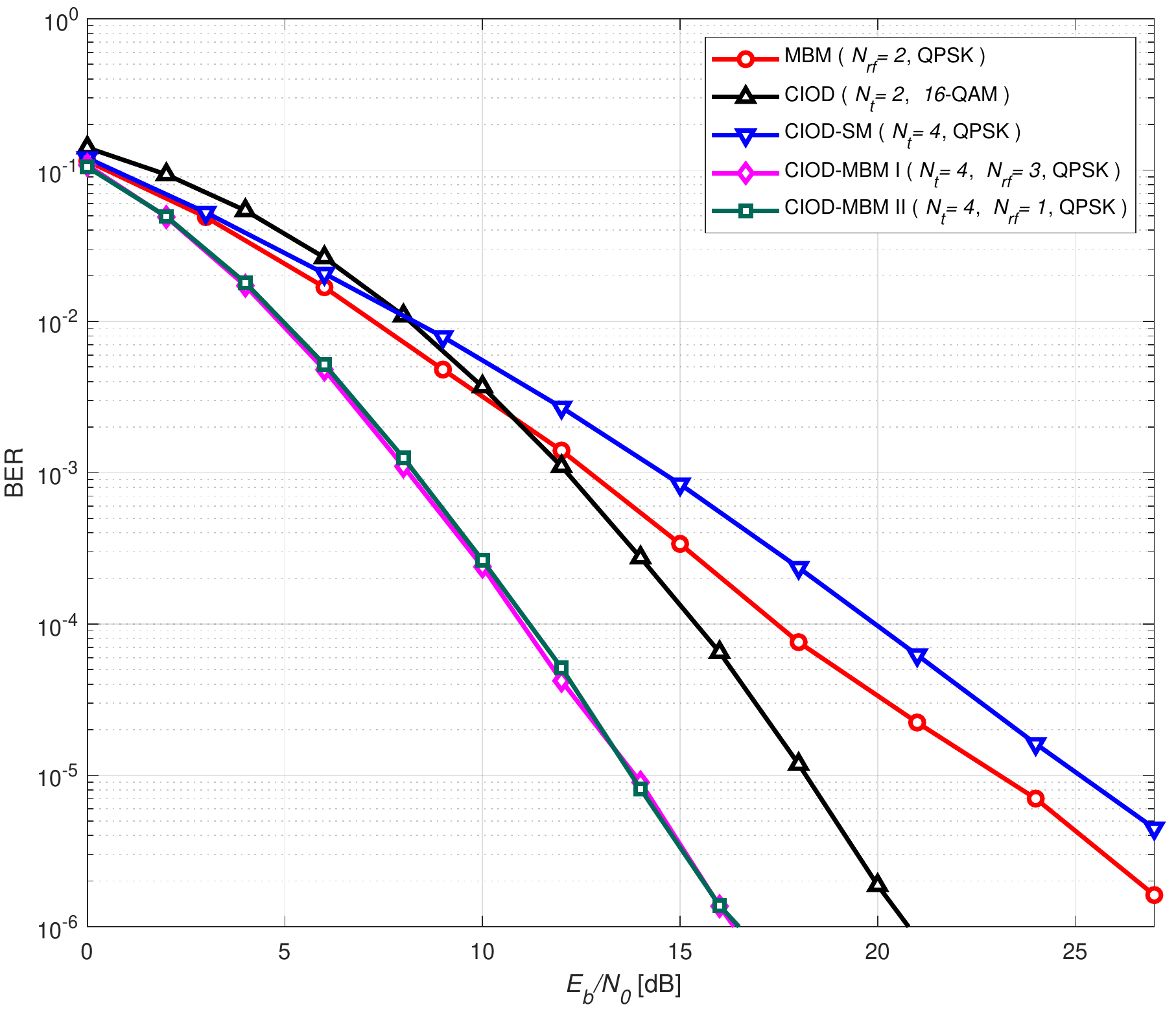}}
		\vspace*{-0.0cm}\caption{The BER of the proposed CIOD-MBM schemes
			compared to the reference systems with $N_r= 2$ receive antennas for about $\eta=4$ bit/s/Hz.}\vspace*{-0.5cm}
	\end{center}
\end{figure}

	\bibliographystyle{IEEEtran}
	\bibliography{IEEEabrv,bib_2019}

\end{document}